\documentclass[aps,prb,twocolumn,superscriptaddress]{revtex4-2}
\usepackage{graphicx}
\usepackage{amssymb}
\usepackage{amsmath}
\usepackage{multirow}

\begin{document}
\title{A$_1$--A$_2$ splitting in pure $^3$He in nematic aerogel}
\author{V.\,V.\,Dmitriev}
\affiliation{P.L.~Kapitza Institute for Physical Problems of RAS, 119334 Moscow, Russia}
\author{M.\,S.\,Kutuzov}
\affiliation{Metallurg Engineering Ltd., 11415 Tallinn, Estonia}
\author{D.\,V.\,Petrova}
\affiliation{P.L.~Kapitza Institute for Physical Problems of RAS, 119334 Moscow, Russia}
\affiliation{HSE University, 101000 Moscow, Russia}
\author{A.\,A.\,Soldatov}
\email{soldatov\underline{ }a@kapitza.ras.ru}
\affiliation{P.L.~Kapitza Institute for Physical Problems of RAS, 119334 Moscow, Russia}
\author{A.\,N.\,Yudin}
\affiliation{P.L.~Kapitza Institute for Physical Problems of RAS, 119334 Moscow, Russia}
\affiliation{HSE University, 101000 Moscow, Russia}

\date{\today}

\begin{abstract}
Here, we present the results of vibrating wire experiments in pure $^3$He (without $^4$He coverage) in nematic aerogel. We investigated the dependence of splitting of the superfluid transition temperature of $^3$He in aerogel on magnetic field. In addition to our previous work, we used a wider range of magnetic fields (up to 31\,kOe) and managed to detect both the ``upper'' and ``lower'' superfluid transition temperatures. The solid paramagnetic $^3$He layer on the aerogel strands activates the magnetic scattering channel. According to theory, it should result in linear splitting at high ($\ge20$\,kOe) fields, while at lower fields the splitting is expected to be nonlinear. We were able to observe this nonlinearity, but we have a discrepancy with theoretical predictions regarding the range of fields where nonlinearity occurs. Possible reasons for this are discussed.
\end{abstract}

\maketitle

\section{Introduction}

Superfluid $^3$He is an example of unconventional (triplet) Cooper pairing with an orbital momentum $L=1$ and a total spin $S=1$. The Ginzburg-Landau free energy and the superfluid transition temperature $T_c$ in bulk superfluid $^3$He are degenerate with respect to spin and orbital momentum projections. In isotropic space up to 18 different superfluid phases are possible to exist having the same transition temperature \cite{Mar}, but in zero magnetic field, depending on temperature and pressure, only two phases (A and B) with the lowest energy are realized \cite{VW}. The A phase belongs to the class of Equal Spin Pairing (ESP) phases (containing Cooper pairs with only $\pm1$ spin projections on a specific direction in spin space) and has an anisotropic superfluid gap with two nodes in the direction of orbital momentum of Cooper pairs. Whereas in the B phase, all spin projections have the same probability and the gap is fully open (isotropic).

Anisotropy of the space may lift the degeneracy, and other phases can be stabilized. For example, in $^3$He in nematic aerogel consisting of nearly parallel strands \cite{asad15} the scattering of $^3$He quasiparticles becomes anisotropic \cite{we15}. It lifts the degeneracy in orbital space and a topologically new ESP phase -- the polar phase -- can be realized \cite{AI,fom14,ik15,fom18,vol18,fom20}. It was discovered in nafen \cite{dmit15} and investigated in detail in many experiments \cite{dm16,zhel16,aut16,dmit18,aut18,dmit19,dmit20,dmit22,elt23}. Both polar and A phases contain $\uparrow\uparrow$ and $\downarrow\downarrow$ spin states of Cooper pairs with equal fractions, where the arrows stand for direction of the magnetic moment.

The degeneracy in spin space is lifted by magnetic field $H$. In bulk $^3$He strong magnetic field splits a transition temperature for different spin components and leads to a formation of new (A$_1$ and A$_2$) phases instead of the A phase. On cooling from normal phase, there are two second-order transitions: ``upper'' transition to the A$_1$ phase at $T=T_{A1}>T_c$ and ``lower'' transition to the A$_2$ phase at $T=T_{A2}<T_c$. The $A_1$ phase has only $\uparrow\uparrow$ pairs and exists in a narrow temperature region expanding linearly with the field \cite{A1,osh,isr84,sag84,koj08,Amb}. The $A_2$ phase has also $\downarrow\downarrow$ pairs, which fraction rapidly increases on cooling, so this phase is continuously transformed to the A phase, where fractions of both spin components become equal to each other.

Similar splitting in magnetic field also takes place in superfluid $^3$He in nematic aerogel \cite{bet}. On cooling, the transition occurs into the so-called $\beta$ phase \cite{sur19_1,sur19_2}, which has the same orbital part of the order parameter as the polar phase, but contains pairs with only $\uparrow\uparrow$ spin states. On further cooling, the transition to the distorted $\beta$ phase, containing also $\downarrow\downarrow$ pairs, is observed. As it was expected, the temperature region of existence of the $\beta$ phase was found to be proportional to $H$ \cite{bet}.

For stabilization of polar and $\beta$ phases it is essential to preplate the strands of nematic aerogel with few atomic layers of $^4$He \cite{dmit15,dmit18,dmit22,bet} to fully remove solid paramagnetic $^3$He from the aerogel surface and ensure a specular scattering for $^3$He quasiparticles \cite{free90,parp92,koj93,st94,mur12}. However, in the absence of $^4$He coverage (pure $^3$He) the strands of aerogel are covered with $\sim$2 atomic layers of paramagnetic solid $^3$He \cite{Sch,Godf,Coll} and the scattering is diffusive. Moreover, the spin is not conserved during the scattering due to a fast exchange between atoms of liquid and solid $^3$He leading to a magnetic scattering channel. Thus, it was established that in pure $^3$He in nematic aerogel, instead of the transition to the polar phase, the transition to the A phase with substantial suppression of a superfluid transition temperature in aerogel $T_{ca}$ takes place \cite{dmit18}. Consequently, the A$_1$-A$_2$ splitting in high magnetic field should occur. In our recent work we conducted experiments in pure $^3$He in nematic aerogel using fields up to 19.4\,kOe to study this splitting \cite{dmit23}. We have found that an ``upper'' transition temperature in nematic aerogel $T_{ca1}$ has a non-linear dependence on applied magnetic field, and further experimental studies in a wider range of magnetic fields are needed to interpret the results.

The A$_1$--A$_2$ splitting in pure $^3$He in aerogel was previously measured only in the case of silica (isotropic) aerogel at the fields up to 8\,kOe with no evidence of the splitting to be found \cite{hal1} and at the fields of 70--150\,kOe \cite{hal2,hal3} with the splitting having a linear field dependence. However, the splitting in silica aerogel at intermediate fields is not measured to the present days. According to Refs.~\cite{ss,bh} the splitting should be modified by the effect of magnetic scattering of $^3$He quasiparticles by polarizable $^3$He spins coating the aerogel strands, so that the splitting becomes non-linear at lower fields and linear at fields $\ge20$\,kOe as the saturation of magnetization of solid $^3$He takes place.

For nematic aerogel we assume a similar phenomena to occur, because no importance of anisotropy of impurities is clearly emphasized within the theory, and the ``upper'' (index 1) and ``lower'' (index 2) transition temperatures in aerogel in a paramagnetic model of the spin exchange are expected to be as follows:
\begin{equation}
\label{tca12}
T_{ca1,2}=T_{ca}\pm\left(\eta_{1,2}-C_{1,2}\frac{\tanh{\alpha H}}{\alpha H}\right) H,
\end{equation}
where $\eta_{1,2}\sim 1\,\mu$K/kOe \cite{Amb,isr84,sag84} are the splitting parameters in the absence of spin exchange, $T_{ca}$ is the superfluid transition temperature of $^3$He in aerogel at $H=0$, $\alpha=\gamma\hbar/(2kT_{ca})$, $\gamma$ is the gyromagnetic ratio, $k$ is the Boltzmann constant, and the spin-exchange parameters $C_{1,2}\sim 1\,\mu$K/kOe \cite{bh} depend on a superfluid coherence length, on a mean free path of $^3$He quasiparticles in aerogel, and on impurity scattering parameters.

In present paper, we measure the A$_1$--A$_2$ splitting in pure $^3$He in nematic aerogel in a wider range of magnetic fields (up to 31\,kOe) and infer the previously observed nonlinearity. In particular, in higher fields the ``lower'' superfluid transition becomes detectable which was not the case at lower fields ($\le19.4$\,kOe) \cite{dmit23}. We see that a magnetic field dependence of the splitting becomes linear much earlier than expected from the theoretical predictions \cite{ss}.

\section{Samples and methods}

Experiments are carried out in pure $^3$He (without $^4$He coverage) using the same setup and the same sample of nematic aerogel as described in Ref.~\cite{dmit23} at a pressure of 15.4\,bar in magnetic fields 1.1 and 19.4--31\,kOe. For the purpose of getting higher magnetic fields compared to our previous work \cite{dmit23}, a new compact superconducting solenoid was designed and wound. The necessary temperatures are obtained by a nuclear demagnetization cryostat and measured using a quartz tuning fork calibrated as described in Ref.~\cite{bet}. The fork is placed in a low-field region to have the B phase around it during the measurements.

We use a vibrating wire (VW) resonator technique with a mullite nematic aerogel sample glued to it, so that the aerogel strands are along the oscillatory motion of the wire. The 240\,$\mu$m NbTi wire is bent into a shape of an arch with height of 10\,mm and distance between legs of 4\,mm. The cuboid sample has sizes $\approx2\times3$\,mm across the strands and 2.6\,mm along the strands, a porosity of 95.2\%, a strand diameter of $\leq14$\,nm, and an average distance between strands of 60\,nm. The wire is mounted on the top of a cylindrical tube of the experimental chamber, made of Stycast-1266 epoxy resin, surrounded by a superconducting solenoid (see the sketch in Ref.~\cite{bet}), so that the sample is located at the magnetic field maximum (with homogeneity of 0.7\% at distances $\pm3$\,mm).

A standard measurement procedure of the aerogel VW resonator is used \cite{CHH}. An alternating current is passed through the wire with amplitude $I_0=0.137\div4$\,mA in accordance with $H$ value and is set to keep the oscillation amplitude field-independent, $HI_0={const}$, where $I_0=0.137$\,mA at $H=31$\,kOe. In magnetic field a Lorentz force acts on a crossbar of the wire leading to periodic oscillations. Motions of the wire in magnetic field, in turn, generate a Faraday voltage which is amplified by a room-temperature step-up transformer {1:300} and measured with a lock-in amplifier by sweeping the frequency. In-phase (dispersion) and quadrature (absorption) signals are joint-fitted to Lorentz curves. At $T\sim1$\,K in vacuum the VW has the resonance frequency of 621\,Hz and the full width at half-maximum (FWHM) of 0.3\,Hz. In liquid $^3$He the wire maximum velocity in the used temperature range did not exceed 0.2\,mm/s. In a given field, measurements with a few times smaller excitation currents were also done and showed the same results.

\begin{figure}[]
\includegraphics[width=\columnwidth]{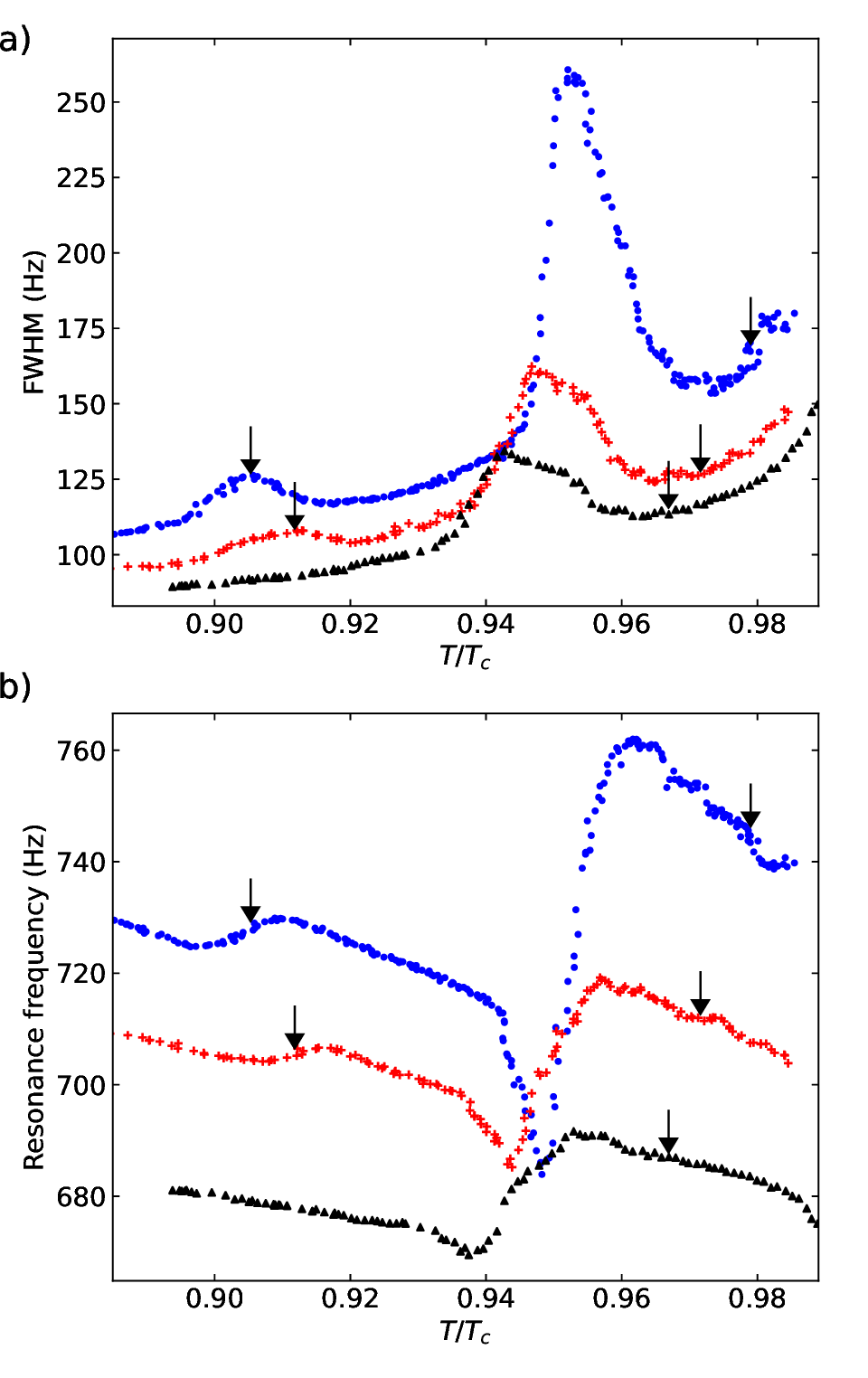}
\caption{Temperature dependence of the resonance width (a) and frequency (b) of the main mode of the aerogel VW in magnetic fields of 19.4\,kOe (triangles), 23\,kOe (crosses), and 31\,kOe (circles). For clarity of view, the triangles and crosses in the panel (b) are up-shifted by 110\,Hz and 74\,Hz respectively. Arrows indicate $T=T_{ca1}$ and $T=T_{ca2}$ determined as described in the text. The x axis represents the temperature normalized to the superfluid transition temperature in bulk $^3$He $T_c=2.083$\,mK.}
\label{f1}
\end{figure}

\begin{figure}[]
\includegraphics[width=\columnwidth]{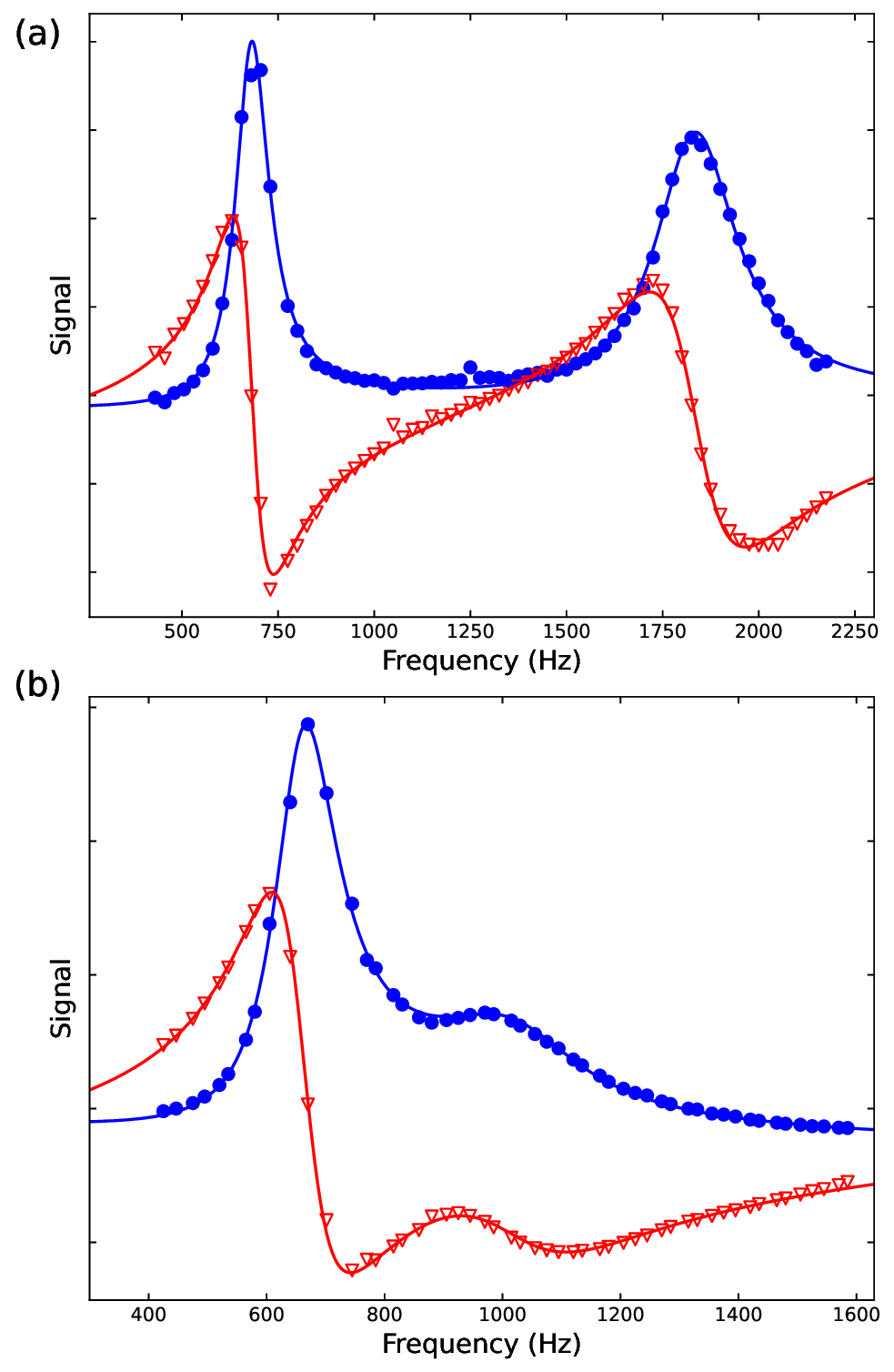}
\caption{Absorption (circles) and dispersion (triangles) signals from the aerogel VW at $T=0.86T_c$ (a) and at $T=0.94T_c$ (b) in magnetic field of 23\,kOe. Solid lines is a result of a joint fit of the data with a sum of two Lorentz curves. The windows on the data are due to removal of a parasitic signal.}
\label{f2}
\end{figure}

\begin{figure}[]
\includegraphics[width=\columnwidth]{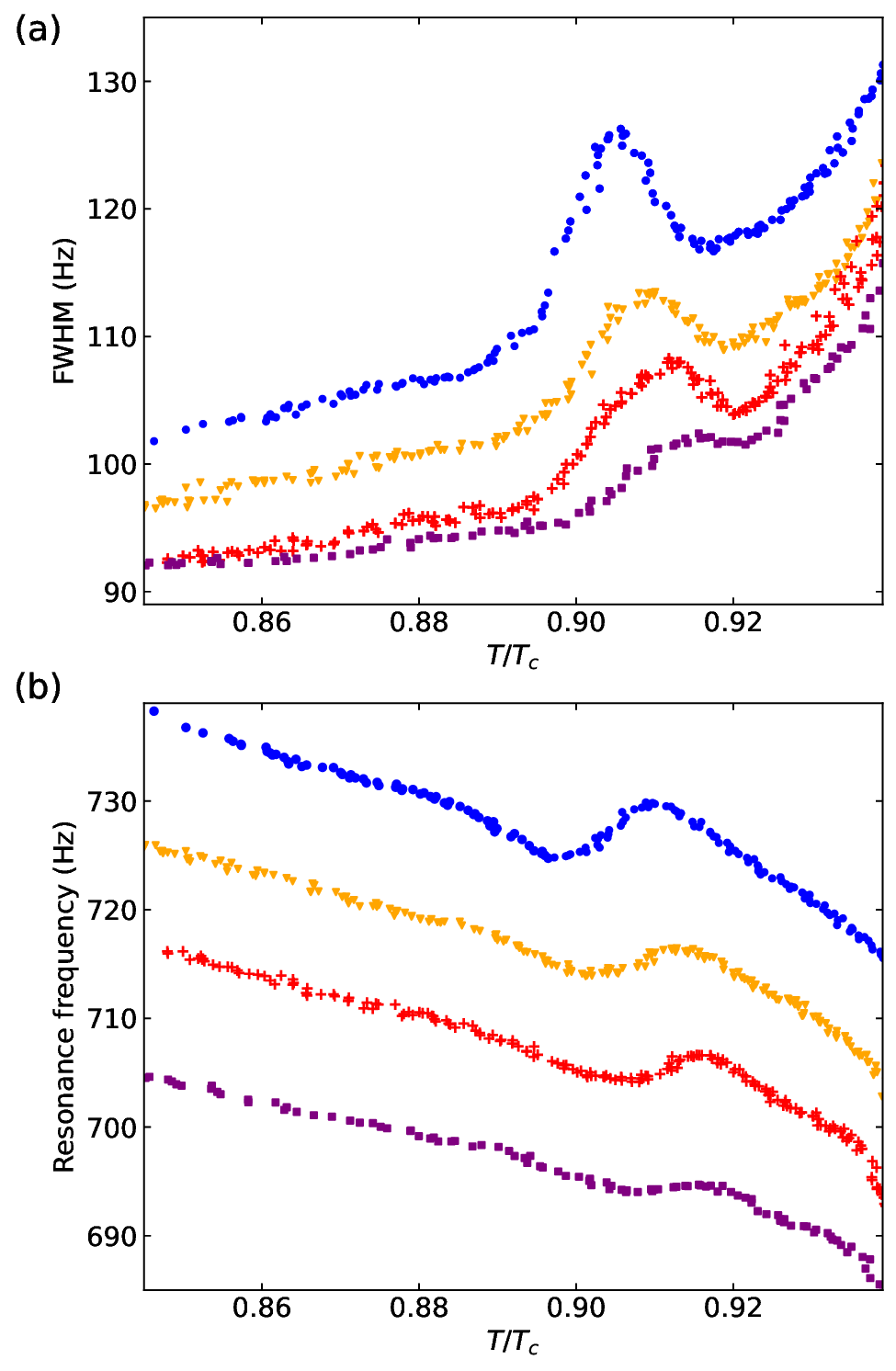}
\caption{Temperature dependence of the resonance width (a) and frequency (b) of the main mode of the aerogel VW in magnetic fields of 21\,kOe (squares), 23\,kOe (crosses), 27\,kOe (triangles), and 31\,kOe (circles) at a temperature region, where a ``lower'' superfluid transition is expected. The triangles, crosses, and squares in the panel (b) are up-shifted by 41, 74, and 90\,Hz correspondingly for clarity of view.}
\label{f3}
\end{figure}

\begin{figure}[]
\includegraphics[width=\columnwidth]{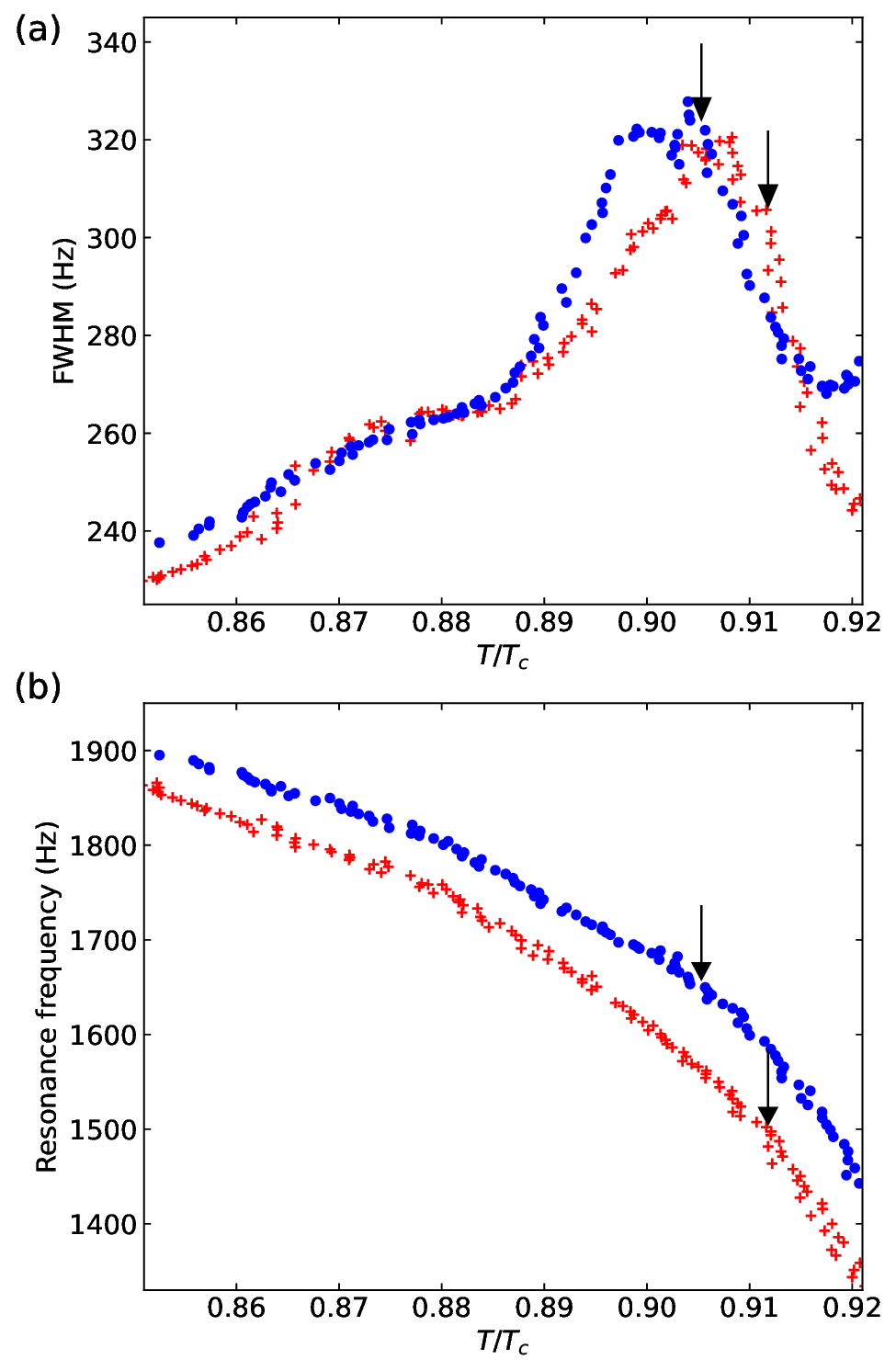}
\caption{Temperature dependence of the resonance width (a) and frequency (b) of the second mode of the aerogel VW in magnetic fields of 23\,kOe (crosses) and 31\,kOe (circles). Arrows indicate $T=T_{ca2}$ determined from the temperature dependence of the width of the main mode (see text).}
\label{f4}
\end{figure}

\begin{figure}[]
\includegraphics[width=\columnwidth]{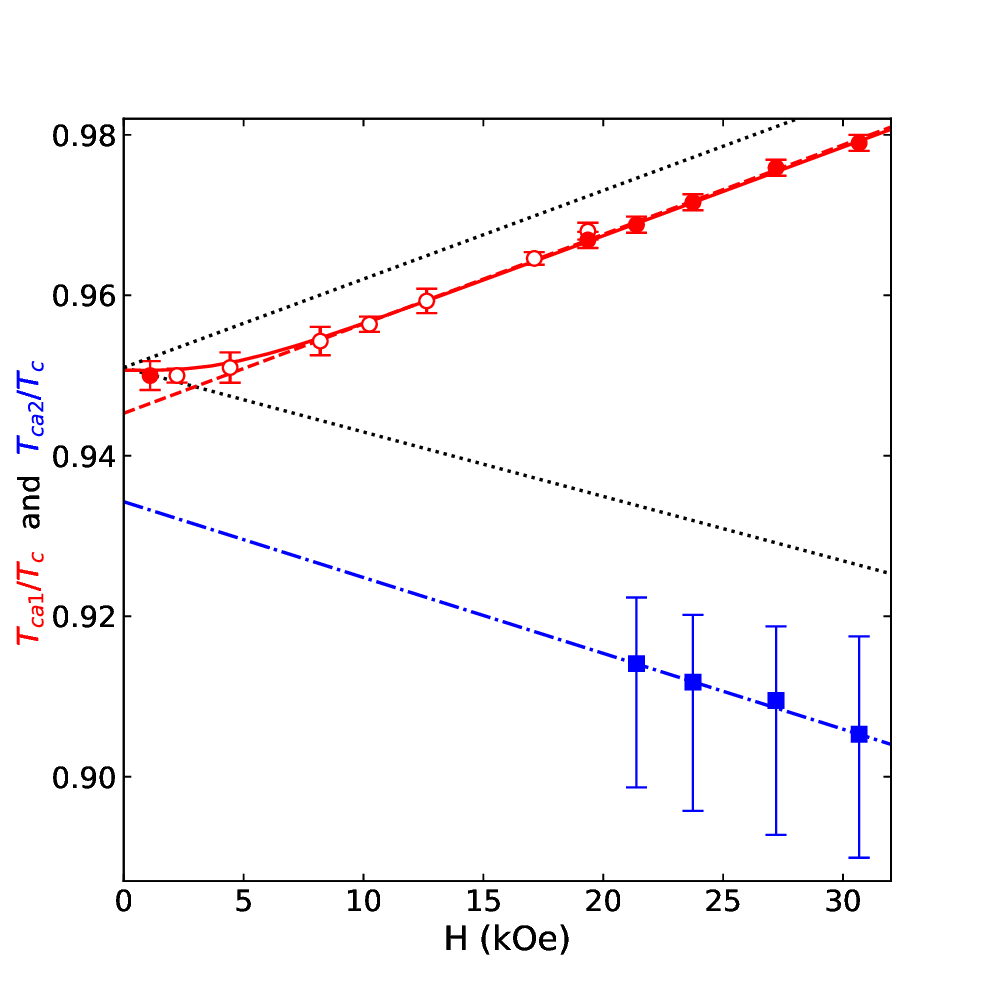}
\caption{The A$_1$--A$_2$ splitting of a superfluid transition temperature of pure $^3$He in nematic aerogel in magnetic field. The circles correspond to the ``upper'' superfluid transition temperature $T_{ca1}$ (open circles are taken from Ref.~\cite{dmit23}), the squares correspond to the ``lower'' superfluid transition temperature $T_{ca2}$. The solid line is the best fit with Eq.~\eqref{tca12}, the dashed line is a linear extrapolation of the high-field $T_{ca1}$ data to a zero field, and the dash-dotted line is a linear fit of the $T_{ca2}$ data. Dotted lines correspond to the A$_1$--A$_2$ splitting in bulk $^3$He scaled to a superfluid transition temperature of $0.95T_c$ \cite{isr84}. The error bars on the $T_{ca2}$ data show a characteristic peak size on the background width of VW resonance in Fig.~\ref{f3}(a) indicating, presumably, the beginning and the end of the A$_1$-A$_2$ transition on cooling.}
\label{f5}
\end{figure}

From the previous VW experiments in pure $^3$He with the same aerogel sample \cite{dmit23} it was found that $\left. T_{ca1}\right|_{H=0}\equiv T_{ca}\approx0.95\,T_c$, where the superfluid transition temperature in bulk $^3$He $T_c=2.083$\,mK, and below $T_{ca1}$ the second oscillating mode in aerogel is excited which is coupled near $T_{ca1}$ to the main flapping resonance of the aerogel wire itself \cite{vw20}. This mode is an analog of the second sound in the superfluid system \cite{sur22}, and its resonance frequency rapidly increases right after the superfluid transition in aerogel on cooling with the subsequent saturation at approximately few kHz \cite{vw20}. Here, by scanning in a wider frequency range, we investigate the temperature dependence of the parameters of both resonance modes.

\section{Results and Discussion}

Firstly, we measure the temperature dependencies of the resonance width and frequency of the main mechanical resonance mode of the aerogel VW. As example, the results in magnetic fields of 19.4, 23, and 31\,kOe are shown in Fig.~\ref{f1}. The peak-like behavior on the temperature dependence of the width is due to an interaction between the main and second resonance modes. On warming in the superfluid phase of $^3$He in aerogel, the frequency of the second mode decreases and, at some point, it reaches the frequency of the first mode resulting in a sharp increase in the resonance width of the first mode. The interaction is small at lower temperatures, where the second resonance is at higher frequencies (far from the main resonance; Fig.~\ref{f2}(a)), and is large near $T_{ca1}$, where both resonance frequencies are close (the second resonance becomes more indistinguishable; Fig.~\ref{f2}(b)). There is also one curious observation: the higher magnetic field, the greater the interaction with the second mode (the higher the peak on the resonance width; Fig.~\ref{f1}(a)), which was not so clearly manifested in experiments with the same aerogel VW, where a complete $^4$He coverage was used (for stabilization of the polar and beta phases) \cite{bet}. The ``upper'' superfluid transition temperature $T_{ca1}$ is determined from the resonance width versus frequency graph, which is an implicit function of temperature, and is slightly higher than a local minimum in Fig.~\ref{f1}(a). The analysis procedure is described in detail in our recent paper \cite{dmit23}.

Secondly, we have managed to detect an additional feature on the temperature dependence of the resonance characteristics of the main mode at $T/T_c=0.89\div0.93$. In more detail it is presented in Fig.~\ref{f3} for the series of magnetic fields and, as we assume, corresponds to the ``lower'' superfluid transition in aerogel (here, we determine $T_{ca2}$ by the local maximum on the temperature dependence of the width). This feature is not seen in magnetic fields $\le19.4$\,kOe (see Fig.~\ref{f1}(a)), so we were unfortunate in our previous measurements \cite{dmit23}. It is worth noting, that the resonance width of the second mode of the aerogel VW also has a distinctive peak at the same temperature region, while the resonance frequency seems to be consistent with the results of measurements in nematic aerogel using $^4$He coverage \cite{vw20} and has no features (see Fig.~\ref{f4}). In any case we could not extract any reliable information from it.

The results for the A$_1$--A$_2$ splitting in pure $^3$He in nematic aerogel are shown in Fig.~\ref{f5}. The ``lower'' transition temperature $T_{ca2}$ has a linear dependence in applicable magnetic fields. As for the ``upper'' transition temperature $T_{ca1}$, there is a nonlinearity in a low field region. The solid line on the graph is the best fit using Eq.~\eqref{tca12} with $\eta_1$, $C_1$, and $\alpha$ as fit parameters. And according to this fit $\alpha\approx0.215$\,kOe$^{-1}$, which is almost 6 times higher than the theoretical value $\alpha=\gamma\hbar/(2kT_{ca})\approx0.037$\,kOe$^{-1}$ meaning that the observed region of nonlinearity is 6 times smaller than expected. In other words, the extrapolation of the high-field (linear) region of the splitting to $T_{ca1}$ at zero field is a direct measure of the exchange field, which is in our case $\approx4$\,kOe, while according to theory it should be $\approx20$\,kOe \cite{sauls04}.

Previously, having only data at $H\le19.4$\,kOe (open symbols in Fig.~\ref{f5}) was not enough to make a good approximation. Now we see that linear contributions to the splitting (the slopes of dashed and dot-dashed lines in Fig.~\ref{f5}) at higher fields are in a good agreement with that in bulk $^3$He (dotted lines in Fig.~\ref{f5}) and our assumption about a possible change of the order parameter (the A phase acquiring a polar distortion) with a subsequent increase in $T_{ca}$ in magnetic field \cite{dmit23} is not confirmed. However, a rather huge discrepancy of our results with the theory suggests that the paramagnetic model of spin exchange \cite{bh,ss} is not applicable.

Probably, the observed effect is due to fluctuations of an amplitude of the order parameter caused by a magnetic scattering \cite{sur25} (similar to a decrease of $T_{ca}$ caused by correlations in the arrangement of chaotically distributed aerogel impurity centers in zero magnetic field \cite{fom08}). While in high magnetic fields the splitting is linear and these fluctuations should obviously be suppressed, in low magnetic fields (much smaller than its spin saturation value) fluctuations can have more influence on the splitting rather than a spin-exchange scattering channel, which is known to be proportional to the average spin \cite{bh,ss}. Besides, we have a gap of $\approx0.01T_c$ on the splitting at zero field, presumably due to a systematic error in determination of $T_{ca1}$ and $T_{ca2}$, which are also expected to be suppressed by temperature fluctuations of the order parameter caused by chaotically distributed aerogel strands \cite{fom08}.

Further experiments using an isotropic (silica) aerogel sample may shed light on the role of magnetic scattering effects in superfluid $^3$He in aerogel.

\section{Conclusions}

In present paper by means of an aerogel VW resonator in pure $^3$He (without $^4$He coverage) we have measured the field dependence of an ``upper'' superfluid transition temperature of $^3$He in nematic aerogel in expanded magnetic fields and have established that a region of nonlinearity of this dependence is about 6 times smaller than expected from theory. Besides, we have managed to detect a ``lower'' superfluid transition temperature of $^3$He in nematic aerogel in higher magnetic fields. Linear sections of the A$_1$--A$_2$ splitting of pure $^3$He in nematic aerogel are found to have basically the same slopes as expected in bulk $^3$He. Further experimental and theoretical work is needed for interpretation of the results.

\begin{acknowledgments}
The aerogel sample was made and provided by M.S.K. The experiments were carried out by V.V.D., D.V.P., A.A.S., and A.N.Y. We are grateful to I.A.~Fomin and E.V.~Surovtsev for useful discussions and comments.
\end{acknowledgments}

\end{document}